\documentclass{article}
\usepackage[letterpaper]{geometry}
\usepackage{setspace,graphicx,overcite}

\AtBeginDocument{\doublespacing}

\begin{document}

\vspace{7.5 mm}

Modeling low energy sputtering of hexagonal boron nitride by xenon ions\\

John T. Yim

\it Department of Aerospace Engineering, University of Michigan, Ann Arbor, Michigan 48109 \rm

Michael L. Falk

\it Department of Materials Science and Engineering, University of Michigan, Ann Arbor, Michigan 48109 \rm

Iain D. Boyd

\it Department of Aerospace Engineering, University of Michigan, Ann Arbor, Michigan 48109 \rm\\

(Received\\

\begin{abstract}
The sputtering of hexagonal boron nitride due to low energy xenon ion bombardments occurs in various applications including fabrication of cubic boron nitride and erosion of Hall thruster channel walls.  At low ion energies, accurate experimental characterization of sputter yields increases in difficulty due to the low yields involved.  A molecular dynamics model is employed to simulate the sputtering process and to calculate sputter yields for ion energies ranging from 10~eV to 350~eV.  The results are compared to experimental data and a semi-empirical expression developed by Bohdansky is found to adequately describe the simulation data.  Surface temperature effects are also investigated, and the sputter yield at 850~K is approximately twice that at 423~K.
\end{abstract}

\newpage

\section{INTRODUCTION}
Boron nitride (BN) is of high interest because it is a chemically inert high temperature material that exhibits high thermal conductivity and a low dielectric constant.\cite{vel:matscieb,lipp:jeurocs}  Two forms of BN are of particular interest.  The extreme hardness of the cubic allotrope of BN (c-BN) makes it a promising material for protective or abrasive coatings, and there is much active research to improve fabrication techniques.\cite{vel:matscieb,mirkarimi:matscie}  The hexagonal allotrope of BN (h-BN) has characteristics similar to graphite, just as c-BN has characteristics similar to diamond.  This form of BN is easily synthesized and machinable.  It is used for insulators in space electric propulsion thrusters as well as substrates for electronic devices, crucibles and molds for high-temperature applications, and as a solid lubricant.\cite{lipp:jeurocs}

Characterization of BN sputtering due to ion bombardment is important for both c-BN and h-BN.  Ion bombardment can play a major role in the synthesis of c-BN thin films through ion-assisted deposition techniques.\cite{mirkarimi:matscie}  The main motivation for the present work, however, comes from the sputtering of the h-BN insulator walls of Hall thrusters, a form of plasma-based space propulsion.  Erosion of the discharge channel walls due to ion impingement is one of the main life-limiting factors for the stationary plasma thruster (SPT) type of Hall thrusters.  Space missions utilizing these thrusters require operational times that reach into the thousands to tens of thousands of hours.  A purely experimental approach is very time-consuming and costly, and so modeling efforts to simulate the wall erosion are an important means of validating the thruster lifetimes.  The results of these erosion models tend to be sensitive to the sputter yields chosen, particularly at low ion energies ($<$~300~eV), and so an accurate characterization of the sputter yields in this energy range is required.

There are continuing efforts to experimentally gather the sputter yields of h-BN at low ion energies.  These incorporate sensitive measurement techniques, such as quartz crystal microbalance (QCM) and cavity ring-down spectroscopy (CRDS), to measure sputter yields for ion energies as low as 100~eV.\cite{yalin:iepc07, rubin:iepc07}  A modeling approach complimentary to these experimental efforts is useful for comparing results and extrapolating to lower ion energies.  We present such a model here.

We use a molecular dynamics (MD) model to simulate the sputtering of h-BN due to low energy xenon ions.  The MD approach is chosen as it provides a method that is as close to a first-principles approach as possible while still maintaining tractable solutions.  The MD method has been used successfully in a wide variety of areas including sputtering simulations.\cite{zhou:nuclinst05,moore:nimpb04,kress:jvacsta99,kubota:jap98}  It has been used before for nitride sputtering simulations, however only purely repulsive potentials were used and to compensate for that, a surface energy barrier was incorporated.\cite{eltekov:radeff95,promokhov:vac00,elovikov:radeff03}  The method presented in this work incorporates a full MD method without any additional assumptions or caveats to model the sputtering of h-BN as accurately as possible.  Another method that is popular for sputtering studies uses a Monte Carlo approach based on the binary collision approximation (BCA), such as the code TRIM.\cite{biersack:applphysa84,eckstein:applphysa85,chen:ieee98}  Though it provides quick results that often match well with experimental data, the BCA assumption breaks down at low ion energies where multi-body collisions become important.\cite{sigmund:physrev69,urbassek:nimb97}  In addition, sputter yields calculated using BCA methods depend on a surface binding energy parameter which is not precisely known for a target material and is often fitted to experimental data.  

Details of the MD model are described in the next section.  Calculated sputter yields and comparisons to experimental and analytical results are provided in the results section.  The effects of ion energy, ion incidence angle, and material temperature are investigated.  A summary and conclusions are presented in the final section.

\section{MODEL}
The interatomic potential for the interaction among the boron and nitrogen atoms is taken from the one presented by Albe {\it et al.}\cite{albe:radeff97, albe:compmatsci98}  Albe's potential function is a modified form of Tersoff's bond-order potential, which incorporates environmental influences, such as the coordination number, bond lengths, and bond angles, to determine the strength of the interatomic bonds.\cite{tersoff:physrevlett86, tersoff:physrevb88}  The Albe potential is fitted to a number of two and three-body boron and nitrogen configurations whose structural properties are known, while also maintaining good comparisons to various phases of bulk BN.\cite{albe:radeff97}  These characteristics make the potential function a good candidate for simulating the sputtering process since it involves modeling various states ranging from the ordered lattice structure to sputtered compounds.

Two modifications are made from the original Albe potential function.  First, the cutoff function, that limits the interaction range, is changed from the sine-based one to the following\cite{los:physrevb03}
\begin{equation}
  f_c\left(r\right) = 
  \left\{ \begin{array}{ll}
    1, & r \leq R - D \\
    \exp \left( \alpha \frac{x^3}{x^3 - 1} \right), & R - D < r < R + D \\
    0, & r \geq R + D \\
  \end{array} \right.
  \label{exp_cutoff}
\end{equation}
\begin{equation}
  x = \frac{r - (R - D)}{2D}
  \label{exp_cutoffB}
\end{equation}
where $r$ is the distance between the two atoms of interest and $R$ and $D$ set the radii of the cutoff shell.  The first and second derivatives of the cutoff function given by Eqs.~(\ref{exp_cutoff}) and (\ref{exp_cutoffB}) are equal to zero at both ends of the cutoff function range; this provides for a smoother transition than the original sine-based cutoff function.  The magnitude of the local extremum of the first derivative is minimized when the coefficient $\alpha$ is set to $3$.

The second change that is made to the Albe potential function is a reduction of the sensitivity of the potential to bond angles involving boron-boron bonds.  The bond angle term in the potential function is given by\cite{albe:radeff97}
\begin{equation}
  g(\theta_{ijk}) = 1 + \frac{c^2}{d^2} - \frac{c^2}{d^2 + (h - \cos(\theta_{ijk}))^2}
  \label{g}
\end{equation}
where $\theta_{ijk}$ is the bond angle and $c$, $d$, and $h$ are fitting coefficients.  When the $i$ and $j$ atoms are both boron, $c$ and $d$ are set to 0.52629 and 0.001587, respectively, in the original function.  The low value of $d$ makes the last term of Eq.~(\ref{g}) very sensitive around the critical angle determined by $h$.  Keeping these values would require a prohibitively small time step for the simulation in order to resolve the potential and conserve energy properly.  Therefore, the values of $c$ and $d$ are increased to 3.316257 and 0.01 respectively.  This retains the relative ratios of $c^2$ and $d^2$ for Eq.~(\ref{g}), but reduces the sensitivity around the critical angle such that a reasonable time step can be used for the calculations.  Normally such changes to the potential would be tested by observing the effects on the elastic and other properties of the material.  However, as B-B bonds do not appear in the BN structure, there is no effect on the BN properties. 

As the xenon ions are assumed to be neutralized by an electron from the surface before impact, electrostatic effects do not need to be considered and the xenon particles are modeled as neutral atoms.\cite{wilhelm:austjphys85}  Also, since the van der Waals attraction of the xenon with the boron and nitrogen atoms is much weaker than the covalent bonding in the boron nitride, a purely repulsive potential is acceptable.\cite{kalyanasundaram:actamat06}  The Moli\`{e}re potential function, a common choice for ion impact studies, is chosen for the xenon ion interactions in this work.

\begin{figure}[h]
 \begin{center}
  \includegraphics{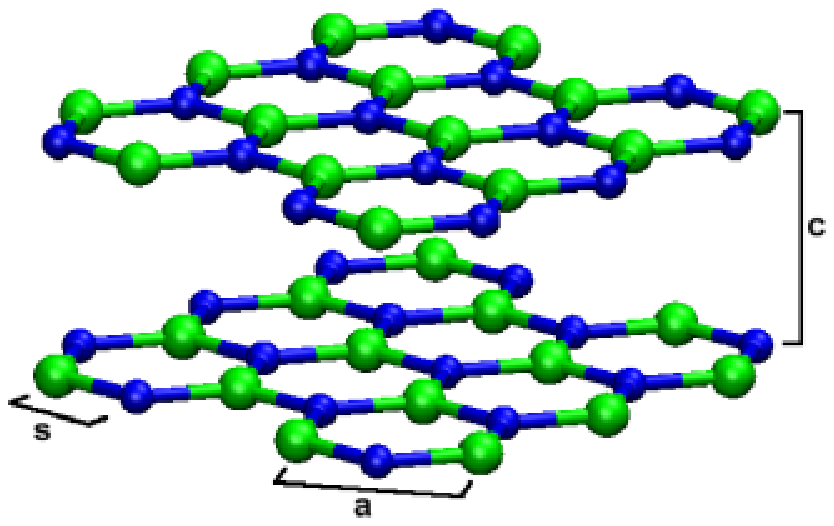}
  \caption{Sample view of an h-BN lattice.}
  \label{hBN}
 \end{center}
\end{figure}

The boron nitride surface is modeled as a block with periodic boundary conditions in the two lateral directions.  The bottom layer of atoms is fixed to prevent translation of the block.  Hexagonal boron nitride is structurally similar to graphite.  Alternating boron and nitrogen atoms are placed in hexagonal lattices which are arranged as a series of sheets.  Equilibrium dimensions\cite{albe:radeff97} and a sample view are shown in Table~\ref{hBN_table} and Fig.~\ref{hBN}, respectively.  Three domain sizes are used to model the BN surface.  Due to the periodic boundary conditions in the lateral directions, a sufficiently large domain size is necessary to prevent system size effects.  The smallest domain, used for ion energies at 100 eV or below, consists of 24 sheets that are 18 hexagons wide and 10 hexagons high, resulting in a 7.8~nm $\times$ 7.8~nm $\times$ 2.5~nm BN block with 18144 atoms.  A larger domain, used for ion energies between 100 and 250~eV, consists of 32 sheets of 24 $\times$ 12 hexagons, or a 10.4~nm $\times$ 10.4~nm $\times$ 3.0~nm block with 38400 atoms.  The largest domain, for ion energies at 250~eV and above, consists of 40 sheets of 30 $\times$ 12 hexagons, or a 13.0~nm $\times$ 13.0~nm $\times$ 3.0~nm block with 60000 atoms.  Domain size independence is established for the two smaller domains for the ion energies they are used for.  Domain size independence is not verified for the largest domain used for 250~eV and 350~eV ions due to the computational cost required for larger domains.  Typically, 30 to 45 minutes of wall clock time are required per ion impact for the lowest domain size on a dual-processor machine.  For the largest domain size used, roughly 3 to 4 hours are required per ion on the same computing system.

The temperature of the BN system is regulated through a Berendsen thermostat which is applied to the two layers of atoms right above the immobile layer.  The rest of the BN system is eventually regulated through conduction.  The Berendsen thermostat rescales the velocities through a factor\cite{berendsen:jchemphys84}
\begin{equation}
  \lambda = \sqrt{1 + \frac{\Delta t}{\tau}\left(\frac{T_0}{T} - 1\right)}
  \label{berendsen}
\end{equation}
where $\Delta t$ is the time step, $\tau$ is a relaxation time parameter, $T$ is the current temperature, and $T_0$ is the bath temperature.  The ratio of $\tau$ to the time step $\Delta t$ is kept around 0.0025.  The target equilibrium temperature, $T_0$, is set to 150$^\circ$C, as that is the temperature reported in several of the experimental tests.\cite{garnier:jvac99,yalin:jpc06,rubin:iepc07}  A sub-relaxation technique is employed to gauge the macroscopic temperature of the system.\cite{sun:jtht05}  The initial temperature is set through the initial particle velocities which are chosen randomly from a Maxwellian distribution.  A correction is applied to ensure the net velocity of the entire system is zero.  

The leapfrog method is used to integrate the system through time.  Due to the dynamics of the system and the integration scheme chosen, a time step of 0.1~fs is used for the simulations.  The time step is increased after the high energy dynamics of the ion impact have subsided.  This is determined by measuring when the energy of every particle has fallen below a specified threshold value.  To further speed up the calculations, cell lists as well as Verlet lists are employed.\cite{haile:mds}  The code also incorporates OpenMP directives to allow for parallel processing on shared memory systems.

\begin{figure}[h]
 \begin{center}
  \includegraphics{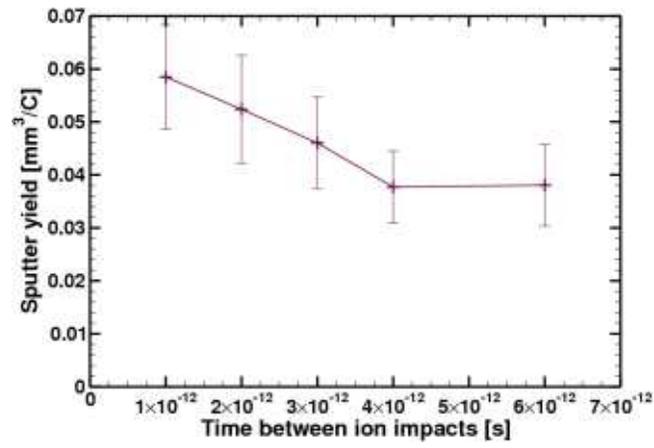}
 \end{center}
 \caption{The calculated BN sputter yields as a function of the time between ion impacts.}
 \label{syflux}
\end{figure}

Ions are inserted into the system one at a time far above the BN surface.  The ions are given an initial trajectory based on a specified incident angle from the surface normal.  The azimuthal angle is chosen randomly to reduce the effects of lattice orientation on the sputtering results.  After each ion insertion, the BN surface is re-equilibrated to the desired temperature and an additional 2~ps are simulated after re-equilibration.  For this work, the vast majority of sputtering events occur within 2~ps after ion insertion; this has also been noted by other sputtering researchers.\cite{zhou:nuclinst05}  Often this results in a total simulated time of 4~ps for each ion.  A parametric study based on the average run time between ion impacts, displayed in Fig.~\ref{syflux}, shows that a 4~ps delay is sufficient for obtaining sputter yield results that are independent of the delay time.

\begin{figure}[h]
 \begin{center}
  \includegraphics{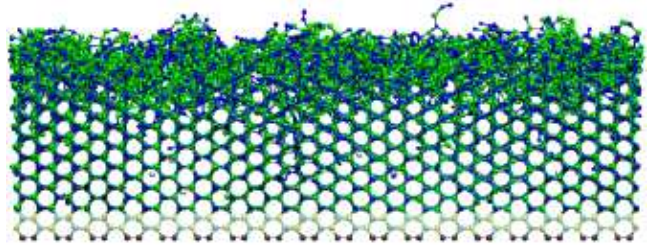}
 \end{center}
 \caption{An end-on view of h-BN showing the amorphous region at the surface after a number of ion impacts have already occurred.  The dark layer at the bottom is the immobile layer and the two lighter layers above it are the thermostat layers.}
 \label{hexlayers}
\end{figure}

Before sputtering statistics are gathered, the system is initialized by bombarding the surface with a number of ion impacts since the sputter yields from a defect-free lattice is different than that from one that has been subjected to ion impacts and implantation.  An amorphous surface layer develops as seen in Fig.~\ref{hexlayers}.  The immobile and thermostat layers are also highlighted in the figure.  After a structural steady state is established, statistics for the sputtering are gathered.  Sputtered particles are detected as they cross a plane a prescribed distance above the surface.  

\section{RESULTS AND DISCUSSION}

\begin{figure}[h]
 \begin{center}
  \includegraphics{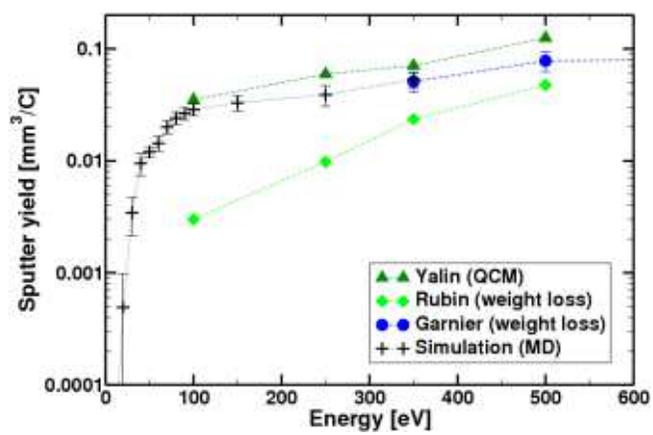}
 \end{center}
 \caption{The total BN sputter yield versus ion energy with an ion incidence angle of 45 degrees on a semi-log scale.}
 \label{sy}
\end{figure}

The total sputter yields are calculated for a range of ion energies, and these are shown in Fig.~\ref{sy}.  In addition to the MD calculated sputter yields, published yields measured using weight loss and QCM techniques are also shown.\cite{garnier:jvac99,yalin:jpc06,rubin:iepc07}  Each set of data are from an ion incidence angle of 45$^\circ$.  The error of the experimental results, if reported, are shown.  For the MD results, the error bars represent the standard deviation of the sputter yield averages.  Typically 300 ion impacts are simulated for the results shown in Fig.~\ref{sy}, but only 200 ion impacts are simulated for the cases at 250~eV and 350~eV due to computational constraints with the largest domain size.  An extended run of 1500 ion impacts is tested for the 50~eV case.

\begin{figure}[h]
 \begin{center}
  \includegraphics{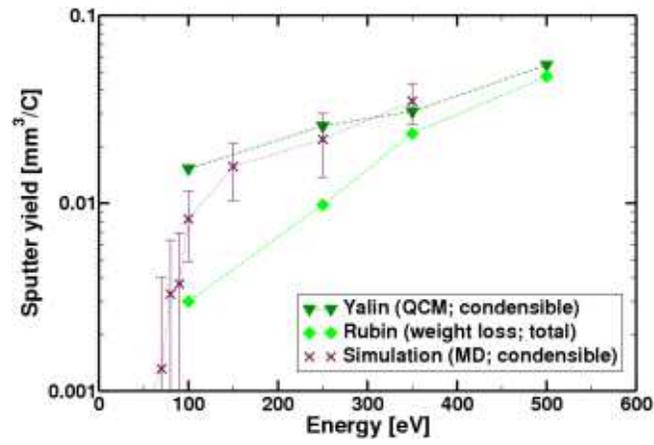}
 \end{center}
 \caption{Comparison of the condensible sputter yields on a semi-log scale.}
 \label{sycond}
\end{figure}

There is a significant amount of scatter among the different sets of experimental data, but overall, the MD results fall within the expected range.  The data from Garnier {\it et al.}\cite{garnier:jvac99} is from a different research group than the other two sets of results.  The grade of BN used is not specified, but they use pyrolytic BN, which has greater than 99\% purity.\cite{lipp:jeurocs}  The QCM results by Yalin {\it et al.}\cite{yalin:jpc06} and the weight loss results by Rubin {\it et al.}\cite{rubin:iepc07} use the same grade of BN---the HBC grade is General Electric's highest purity grade---and similar sample preparation techniques.  The QCM actually only measures compounds that are able to condense on its surface.  In the case of BN, that includes atomic boron and compounds containing boron; the QCM would not register atomic or molecular nitrogen.  In Fig.~\ref{sy}, estimates of the total sputter yields based from the condensible measurements provided by Yalin {\it et al.} are shown.\cite{yalin:jpc06}  Figure~\ref{sycond} displays only the condensible products for both the QCM measurements and the MD simulations.  Though there is a fair amount of agreement between the two sets of condensible data, they are both still higher than the total sputter yields measured using the weight loss method.  The origin of the difference between the various experimental sputter yield results is unclear.  The MD results, however, agree fairly well with the QCM and the Garnier weight loss results.

\begin{figure}[h]
 \begin{center}
  \includegraphics{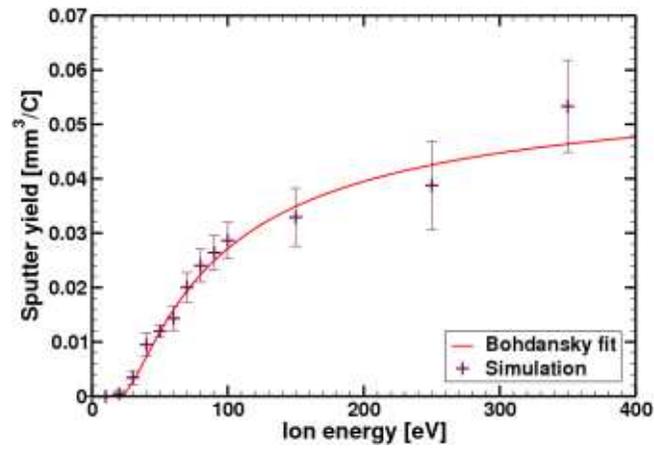}
 \end{center}
 \caption{The Bohdansky fit applied to the MD results.}
 \label{bsyfit}
\end{figure}

To provide an expression for the low energy BN sputter yields, a curve fit is applied to the MD data.  The Bohdansky\cite{bohdansky:nimpr84} formula, a semi-empirical fit based on the work of Sigmund,\cite{sigmund:physrev69} is used.  Yamamura,\cite{yamamura:adndt96} Matsunami,\cite{matsunami:radefflett80} and Zhang\cite{zhang:radeff04} have also based semi-empirical fits on Sigmund's initial work that show good agreement with experimental data at higher energies.  Unfortunately, a few of the assumptions these formulae are derived from break down at very low ion energies.\cite{sigmund:physrev69}  Bohdansky tailors his formula with an emphasis on matching very low energy ion experimental sputtering data.  A least-squares fitting routine provides coefficients for the expression,
\begin{equation}
  Y = \alpha \left[ 1 - \left( \frac{E_{th}}{E} \right)^{2/3} \right] \left[ 1 - \frac{E_{th}}{E} \right]^2
  \label{bsyfiteq}
\end{equation}
where $\alpha = 6.2\times10^{-2} \pm 0.4\times10^{-2}$~mm$^3$/C and $E_{th} = 18.3 \pm 1.1$~eV for ion energies, $E$, also in eV.  Figure~\ref{bsyfit} displays the Bohdansky fit to the MD sputter data.  Equation~(\ref{bsyfiteq}) is for xenon ions with a 45$^\circ$ incidence angle onto a h-BN surface.  The sputter yield likely asymptotically approaches zero instead of sharply cutting off as other models may predict.  An inflection point in the sputter yield energy dependence curve is a notable feature for sputtering at very low ion energies and marks a transition from a sharp decrease to a gradual approach to the threshold.

\section{CONCLUSIONS}

Determining the sputter yields resulting from low energy ion bombardment presents several challenges for both experiment and modeling.  We have applied the molecular dynamics method to simulate the sputtering of a hexagonal boron nitride surface from low energy xenon ion impacts.  An amorphous layer develops at the surface after a number of ions impact the initial lattice structure.  The sputtering characteristics are affected by this amorphous layer and should be taken into account when determining steady state sputter yields.  Sputter yields at impact energies from 10~eV to 350~eV are obtained.  The results compare well to the available experimental data, though there is a certain amount of scatter among the different sets of data.  A curve fit based on the semi-empirical Bohdansky expression, which was explicitly formulated to match to low energy sputter data, is applied to provide an expression for the energy dependence.  Other semi-empirical fits based on the Sigmund formula may not capture the low energy sputter yields as well, since the original Sigmund formulation is based on assumptions that break down at low ion energies.  Surface temperature considerations are also taken into account.  An increase from 423~K to 850~K provides a nearly two-fold increase in the sputter rate.  The MD method is shown to be a viable and useful tool for analysis of low energy ion sputtering.

\section*{ACKNOWLEDGMENTS}

J. Yim and I. Boyd gratefully acknowledge support from NASA Glenn Research Center through grant NCC04GAQ56G.  M. Falk acknowledges the support of NSF under Award\# 0510163.

\newpage

\begin{table}[h]
 \begin{center}
  \caption{Dimension lengths of h-BN.}
  \begin{tabular}{cc}
   \hline
   \hline
   Dimension & Length [\AA] \\
   \hline
   $a$ & 2.496 \\
   $c$ & 3.245 \\
   $s$ & 1.441 \\
   \hline
   \hline
  \end{tabular}
 \label{hBN_table}
 \end{center}
\end{table}

\newpage

\newpage

\newpage

\newpage

\newpage

\newpage

\newpage

\begin{figure}[h]
 \begin{center}
  \includegraphics{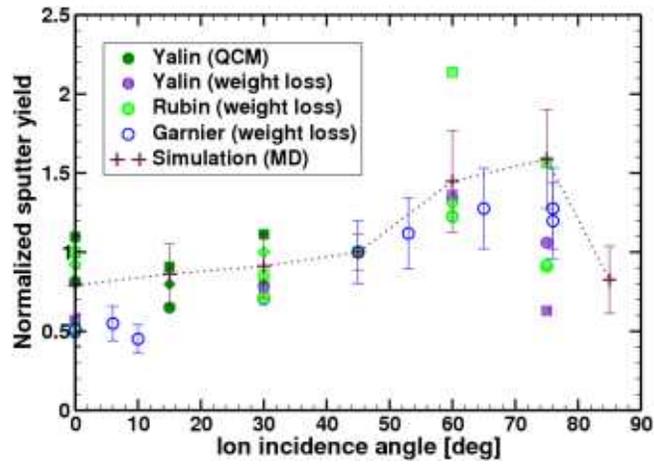}
 \end{center}
 \caption{Normalized sputter yield dependence on incidence angle.}
 \label{syangle}
\end{figure}

\newpage

\begin{figure}[h]
 \begin{center}
  \includegraphics{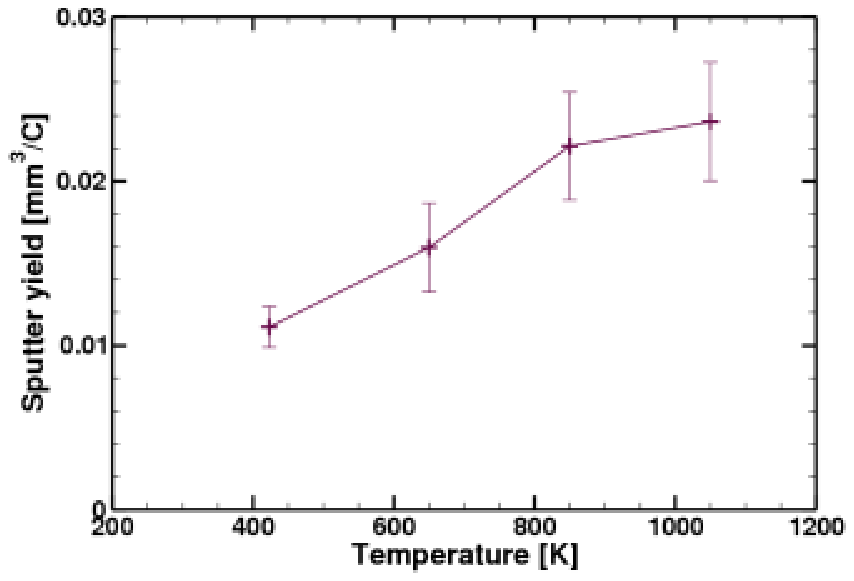}
 \end{center}
 \caption{The sputter yield of BN at various surface temperatures.}
 \label{sytemp}
\end{figure}


\newpage

\begin{thebibliography}{10}

\bibitem{vel:matscieb}
L. Vel, G. Gemazeau, and J. Etourneau, Mat. Sci. Eng. B. {\bf B10}, 149 (1991).

\bibitem{lipp:jeurocs}
A. Lipp, K.~A. Schwetz, and K. Hunold, J. Eur. Ceram. Soc. {\bf 5}, 3 (1989).

\bibitem{mirkarimi:matscie}
P.~B. Mirkarimi, K.~F. McCarty, and D.~L. Medlin, Mater. Sci. Eng. {\bf R21}, 47 (1997).

\bibitem{yalin:iepc07}
A.~P. Yalin, L. Tao, N. Yamamoto, T.~B. Smith, and A.~D. Gallimore, 30th Intl. Elec. Prop. Conf. IEPC 2007-075 (2007).

\bibitem{rubin:iepc07}
B. Rubin, J.~L. Topper, and A.~P. Yalin, 30th Intl. Elec. Prop. Conf. IEPC 2007-074 (2007).

\bibitem{zhou:nuclinst05}
X.~W. Zhou, H.~N.~G. Wadley, and S. Sainathan, Nucl. Instrum. Meth. B. {\bf 234}, 441 (2005).

\bibitem{moore:nimpb04}
M.~C. Moore, N. Kalyanasundaram, J.~B. Freund, and H.~T. Johnson, Nucl. Instr. Meth. B. {\bf 225}, 241 (2004).

\bibitem{kress:jvacsta99}
J.~D. Kress, D.~E. Hanson, A.~F. Voter, C.~L. Liu, X.-Y. Liu, and D.~G. Coronell, J. Vac. Sci. Technol. A. {\bf 17}, 2819 (1999).

\bibitem{kubota:jap98}
N.~A. Kubota, D.~J. Economou, and S.~J. Plimpton, J. Appl. Phys. {\bf 83}, 4055 (1998).

\bibitem{eltekov:radeff95}
V.~A. Eltekov, S.~S. Elovikov, J.~S. Colligon, N.~N. Negrebetskaya, A.~A. Promokhov, and V.~E. Yurasova, Radiat. Eff. Defect. S. {\bf 133}, 107 (1995).

\bibitem{promokhov:vac00}
A.~A. Promokhov, A.~S. Mosunov, S.~S. Elovikov, and V.~E. Yurasova, Vacuum. {\bf 56}, 247 (2000).

\bibitem{elovikov:radeff03}
S.~S. Elovikov, I.~K. Khrustachev, A.~S. Mosunov, and V.~E. Yurasova, Radiat. Eff. Defect. S. {\bf 158}, 573 (2003).

\bibitem{biersack:applphysa84}
J.~P. Biersack and W. Eckstein, Appl. Phys. A. {\bf 34}, 73 (1984).

\bibitem{eckstein:applphysa85}
W. Eckstein and J.~P. Biersack, Appl. Phys. A. {\bf 37}, 95 (1985).

\bibitem{chen:ieee98}
M. Chen, G. Rohrbach, A. Neuffer, K.-L. Barth, and A. Lunk, IEEE T. Plasma Sci. {\bf 26}, 1713 (1998).

\bibitem{sigmund:physrev69}
P. Sigmund, Phys. Rev. {\bf 184}, 383 (1969).

\bibitem{urbassek:nimb97}
H.~M. Urbassek, Nucl. Instrum. Meth. B. {\bf 122}, 427 (1997).

\bibitem{albe:radeff97}
K. Albe, W. M\"{o}ller, and K.-H. Heinig, Radiat. Eff. Defect. S. {\bf 141}, 85 (1997).

\bibitem{albe:compmatsci98}
K. Albe and W. M\"{o}ller, Comp. Mater. Sci. {\bf 10}, 111 (1998).

\bibitem{tersoff:physrevlett86}
J. Tersoff, Phys. Rev. Lett. {\bf 56}, 632 (1986).

\bibitem{tersoff:physrevb88}
J. Tersoff, Phys. Rev. B. {\bf 37}, 6991 (1988).

\bibitem{los:physrevb03}
J.~H. Los and A. Fasolino, Phys. Rev. B. {\bf 68}, 024107 (2003).

\bibitem{wilhelm:austjphys85}
H.~E. Wilhelm, Aust. J. Phys. {\bf 38}, 125 (1985).

\bibitem{kalyanasundaram:actamat06}
N. Kalyanasundaram, M.~C. Moore, J.~B. Freund, and H.~T. Johnson, Acta Mater. {\bf 54}, 483 (2006).

\bibitem{berendsen:jchemphys84}
H.~J.~C. Berendsen, J.~P.~M. Postma, W.~F. van{~}Gunsteren, A. DiNola, and J.~R. Haak, J. Chem. Phys. {\bf 81}, 3684 (1984).

\bibitem{garnier:jvac99}
Y. Garnier, V. Viel, J.-F. Roussel, and J. Bernard, J. Vac. Sci. Technol. A. {\bf 17}, 3246 (1999).

\bibitem{yalin:jpc06}
A.~P. Yalin, V. Surla, C. Farnell, M. Butweiller, and J.~D. Williams, 42nd AIAA/SAE/ASME/ASEE Joint Prop. Conf. AIAA 2006-4338 (2006).

\bibitem{sun:jtht05}
Q. Sun and I.~D. Boyd, J. Thermophys. Heat Tr. {\bf 19}, 329 (2005).

\bibitem{haile:mds}
J.~M. Haile, Molecular Dynamics Simulation : Elementary Methods. Wiley (1992).

\bibitem{bohdansky:nimpr84}
J. Bohdansky, Nucl. Instr. Meth. B. {\bf 2}, 587 (1984).

\bibitem{yamamura:adndt96}
Y. Yamamura and H. Tawara, Atom. Data Nucl. Data. {\bf 62}, 149 (1996).

\bibitem{matsunami:radefflett80}
N. Matsunami, Y. Yamamura, Y. Itikawa, N. Itoh, Y. Kazumata, S. Miyagawa, K. Morita, and R. Shimizu, Radiat. Eff. Lett. {\bf 57}, 15 (1980).

\bibitem{zhang:radeff04}
Z.~L. Zhang and L. Zhang, Radiat. Eff. Defect. S. {\bf 159}, 301 (2004).

\bibitem{yalin:jpc07}
A.~P. Yalin, B. Rubin, S.~R. Domingue, Z. Glueckert, and J.~D. Williams, 43rd AIAA/SAE/ASME/ASEE Joint Prop. Conf. AIAA 2007-5314 (2007).

\bibitem{yamashiro:jap07}
M. Yamashiro, H. Yamada, and S. Hamaguchi, J. Appl. Phys. {\bf 101}, 046108 (2007).

\end{thebibliography}
\end{document}